\documentclass[12pt]{article}
\usepackage{graphicx, color}
\usepackage{cancel}
\usepackage{array,multirow}
\usepackage{amsmath,amssymb,amsfonts}

\newcommand\pubnumber{SNSN-323-63}
\newcommand\pubdate{\today}

\def\grenoble{Ultra Cold Neutron group\\
Laboratoire de physique Subatomique et Cosmologie\\
 38000 Grenoble, FRANCE}

\def\Title#1{\begin{center} {\Large #1 } \end{center}}
\def\Author#1{\begin{center}{ \sc #1} \end{center}}
\def\Address#1{\begin{center}{ \it #1} \end{center}}

\newcommand\pubblock{\rightline{\begin{tabular}{l} \pubnumber\\
         \pubdate  \end{tabular}}}
\newenvironment{Abstract}{\begin{quotation}  }{\end{quotation}}
\newenvironment{Presented}{\begin{quotation} \begin{center} 
             PRESENTED AT\end{center}\bigskip 
      \begin{center}\begin{large}}{\end{large}\end{center} \end{quotation}}

\def\beq{\begin{equation}}
\def\eeq#1{\label{#1}\end{equation}}
\def\eeqn{\end{equation}}

\def\beqa{\begin{eqnarray}}
\def\eeqa#1{\label{#1}\end{eqnarray}}
\def\eeqan{\end{eqnarray}}

\let\bar=\overbar

\def\Dslash{\not{\hbox{\kern-4pt $D$}}}
\def\dslash{\not{\hbox{\kern-2pt $\del$}}}

\def\msb{{\bar{\ssstyle M \kern -1pt S}}}

\def\ecm{~e~\textrm{cm}}
\def\dn{d_{\textrm{n}}}

\def\hg{^{199}}
\def\cs{^{133}}
\def\he{^{3}}

\begin{document}
\begin{titlepage}
\pubblock

\vfill
\Title{The nEDM experiment at the Paul Scherrer Institute}
\vfill
\Author{ Yoann Kermaidic \\ On behalf of the nEDM collaboration} 
\Address{\grenoble}
\vfill
\begin{Abstract}
The quest for a non-zero electric dipole moment (EDM) of simple systems such as the electron, the neutron or atoms / molecules is a powerful way to search for physics beyond the standard model (SM) in particular for new sources of CP violation, complementary to LHC experiments. So far, no EDM signal was observed and the upper limit on the neutron EDM, established by the RAL/Sussex/ILL collaboration, is $|\dn| < 3 \times 10^{-26} \ecm (90\% ~ \textrm{C.L.})$. 
This limits was set with an apparatus using ultra cold neutrons stored in a vessel at room temperature. 
The nEDM collaboration at the Paul Scherrer Institute in Switzerland aims at reaching a sensitivity in the $10^{-27} \ecm$ range soon. 
I will present the current status of the experiment and discuss the prospects for the future.
\end{Abstract}
\vfill
\begin{Presented}
CIPANP2015 conference\\
Denver, USA,  May 19--24, 2015
\end{Presented}
\vfill
\end{titlepage}
\def\thefootnote{\fnsymbol{footnote}}
\setcounter{footnote}{0}

\section{Introduction}

Low energy precision experiments searching in particular for EDMs offer a sensitive approach to probe physics beyond the standard model and already constrain many BSM models.
Measuring a neutron electric dipole moment (nEDM) larger than the SM prediction ($|\dn^{\textrm{SM}}| \approx 10^{-32} - 10^{-31} \ecm$) would reveal new sources of CP violation and provide clues to the baryogenesis puzzle \cite{yk:Pospelov}. \\
\\
Our collaboration is conducting a project to measure the nEDM at the Paul Scherrer Institute's (PSI) ultra cold neutron (UCN) source using a room temperature spectrometer.
We are operating an upgraded version of the RAL/Sussex/ILL (Rutherford Appleton Laboratory, Institut Laue Langevin) apparatus which holds the best nEDM limit : $|\dn| < 3 \times 10^{-26} \ecm (90\% ~ \textrm{C.L.})$ \cite{yk:nEDMlimit}. 
With this apparatus, we aim to slightly improve the nEDM sensitivity within the next year. 
A next generation nEDM apparatus will be built to improve by at least a factor 10 over the present sensitivity.
In section \ref{yk:sec:principle}, I will briefly describe the nEDM measurement principle.
Section \ref{yk:sec:nEDM} is dedicated to the nEDM apparatus description.
Finally, the future n2EDM spectrometer is shortly discussed in section \ref{yk:sec:n2EDM} along with our R\&D program.

\section{Measurement principle}
\label{yk:sec:principle}
The measurement principle is based on the analysis of the Larmor precession frequency of UCN stored in a volume permeated with electric and magnetic fields either parallel or anti-parallel. The Hamiltonian of the system reads:
\begin{eqnarray}
\label{yk:eq:principle}
	h \nu_{\uparrow \uparrow (\uparrow \downarrow)} &=& | 2 ~ \pmb{\mu_{\textrm{n}} \cdot B} \pm 2 ~\pmb{d_{\textrm{n}} \cdot E} | \\
\label{yk:eq:dn}
	h | \nu_{\uparrow \uparrow} - \nu_{\uparrow \downarrow} | &=& 4 | d_{\textrm{n}} | E
\end{eqnarray}
\\
The experiment aims therefore to measure a shift in $\nu$ proportional to the E-field strength.
We use the Ramsey technique of separated oscillating fields \cite{yk:Ramsey} to estimate the neutron Larmor frequency with a precision better than 1 ppm.
It consists of measuring the polarization of neutrons after a defined sequence: i) A $\pi/2$ pulse rotates the longitudinal spin projection in the transverse plane; ii) The spins freely precess in the transverse plane during a time T; iii) A second $\pi/2$ pulse, in phase with the first one is applied.
The sensitivity of this technique reads:
\begin{eqnarray}
\label{yk:eq:ramseysens}
	\sigma_{\nu} = \frac{1}{2\pi T \alpha \sqrt{N}},
\end{eqnarray}
\\
where $\alpha$ is the UCN polarization after $T$ and $N$ is the total number of neutrons.
A nEDM measurement relies on a series of repeated cycles with successive B and E fields configuration.
One should note that Eq. \ref{yk:eq:dn} holds only if $B_{\uparrow \uparrow} = B_{\uparrow \downarrow}$.
This condition requires to make use of highly sensitive atomic magnetometers that have sensitivities better than 0.1 ppm.

\section{The upgraded RAL/Sussex/ILL spectrometer}
\label{yk:sec:nEDM}

The RAL/Sussex/ILL spectrometer was moved from ILL to PSI in 2009 to benefit from the PSI UCN source.
A four cylindrical layers ($\textrm{D}_{\textrm{inner}} = 1.6$ m, $\textrm{L} = 1.1$ m) mu-metal shield having a transverse shielding factor of about 10000 surrounds a 20 $\ell$ storage volume made of two aluminum plates coated with DLC and a deuterated polystyrene hollow cylinder.
The precession chamber is permeated with a highly homogeneous magnetic field of 1 $\mu$T generated by a $\cos \theta$ coil and with a 11 kV/cm electric field.
In order to precisely monitor the magnetic field $B_0$, a $\hg$Hg co-magnetometer that measures the same space and time magnetic field average as the neutron with a typical precision in the 0.05-0.1 ppm range is used \cite{yk:GreenHgM} (Fig. \ref{yk:fig:apparatus}).
See Ref. \cite{yk:nEDMapparatus} for a detailed description of the apparatus, in the following, I will detail the most important upgrades.

\subsection{The PSI UCN source}
\label{yk:sec:UCNSource}

		The UCN source, as operated in 2014, uses the PSI proton beam in a pulsed mode ($I \geq 2.2 \textrm{~mA}, E = 590 \textrm{~MeV}$, mode : 3 s kick every 300 s) to produce neutrons by spallation on a lead target ($\approx$ 8 neutrons/proton) \cite{yk:UCNsourceNIMA}.
		These neutrons are first thermalized in heavy water before being moderated and further converted to UCN in solid deuterium at 5 K \cite{yk:UCNsource}.
		After escaping the solid $D_2$, UCN are guided towards the nEDM spectrometer through tubes coated with NiMo.
		Given the UCN source performance in 2014, one could detect about 6000 UCN on average in the nEDM detector after a 180 s storage time in the precession chamber.
		In 2015 this performance was further improved.
		
\subsection{Neutron detection}

		Before entering into the apparatus, UCN are fully polarized by a 5 T superconducting magnet.
		A set of guiding coils permits to maintain the polarization along the neutron's path and provide an initial polarization of 85 \%.
		To measure the neutron Larmor frequency, we analyse the UCN polarization at the end of the precession time.
		For this purpose, a device for simultaneous spin analysis was built \cite{yk:USSA} based on $^6$Li glass scintillator detectors \cite{yk:NANOSC}.
		The new device treats symmetrically both spin states with two arms, each equipped with an adiabatic spin flipper and an analysing foil made of magnetized iron to select only one spin state.
		This new detection scheme improves our nEDM sensitivity by 18\% per cycle as compared to the original setup, based on a sequential spin analysis \cite{yk:nEDMapparatus}.

\begin{figure}
	\centering
	\includegraphics[height=1.3in]{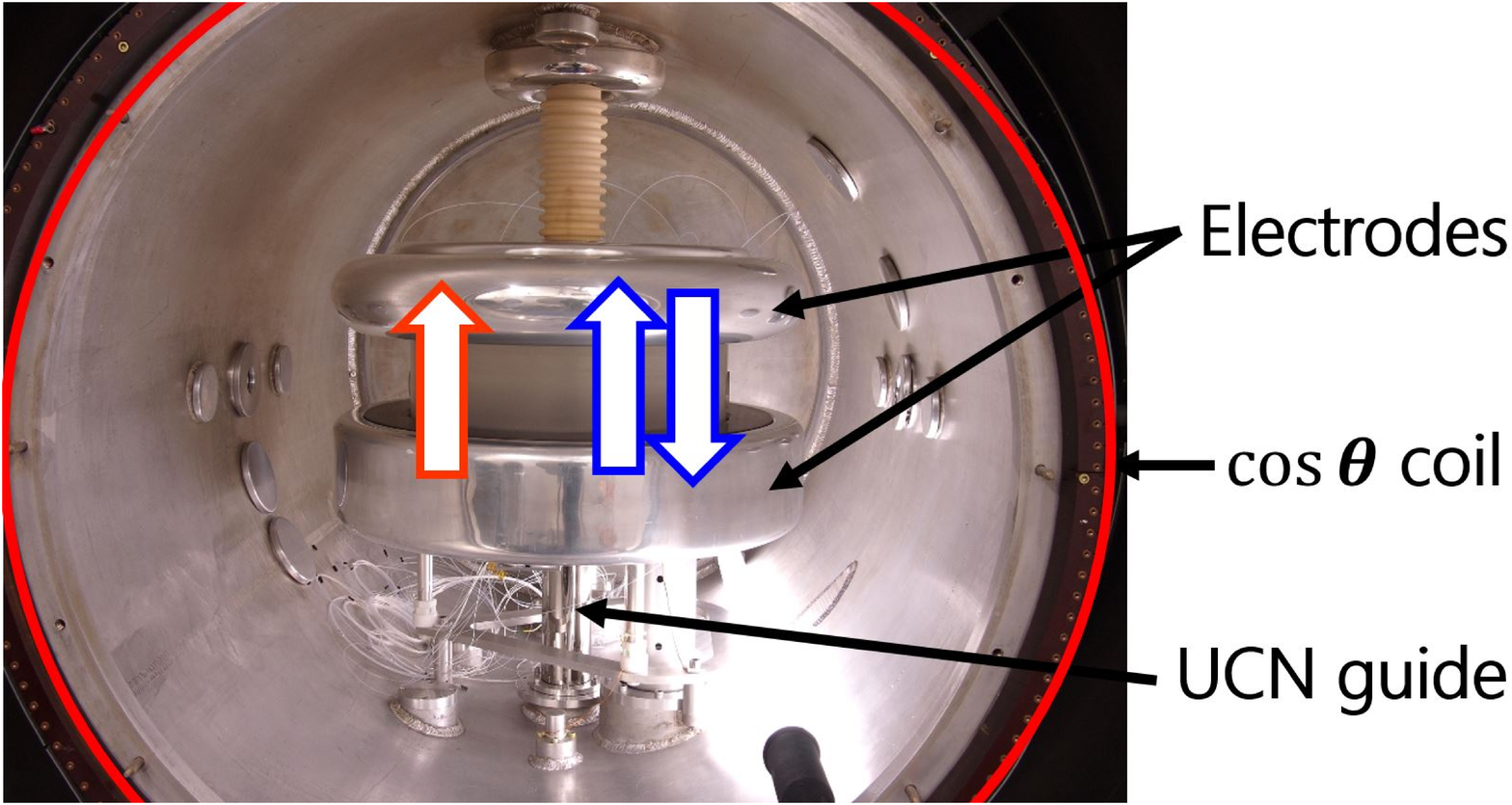}
	\includegraphics[height=1.5in]{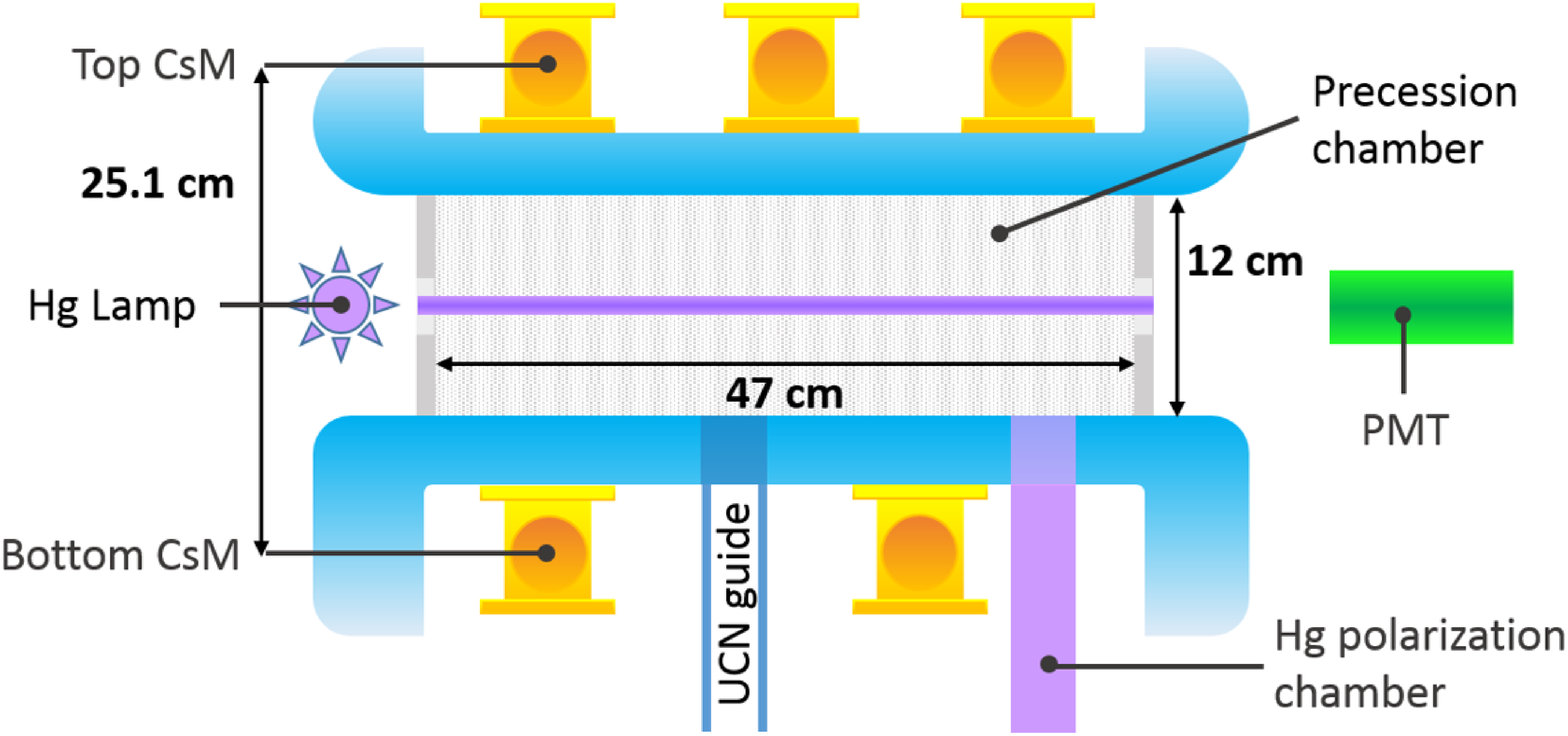}	
\caption{Left : Picture of the nEDM spectrometer inside the vacuum chamber. Red and blue arrows indicate B and E field directions respectively. Right : Vertical cut of the inner nEDM spectrometer. Schematically indicated are the precession chamber, the $\hg$Hg and $\cs$Cs magnetometers.}
\label{yk:fig:apparatus}
\end{figure}
	
\subsection{Atomic magnetometry : $\cs$Cs array}
\label{yk:sec:magnetometry}
	
	An external $\cs$Cs magnetometers array (CsM) was installed (Fig. \ref{yk:fig:apparatus}) and represents one of the main upgrade \cite{yk:CsMarray}.
	It is made of 16 laser-pumped scalar CsM operated in a driven mode: 6 HV-compatible magnetometers placed on top of the precession chamber and the other 10 below.
	This arrangement allows for the measurement of the $B_0$ distribution inside the precession chamber since $B = \sqrt{B_0^2 + B_T^2}\approx B_0$ and gives us information on the averaged vertical gradient $\partial B_0/\partial z$ in the storage volume with a typical uncertainty of 10 pT/cm.
	Whereas the knowledge of $\partial B_0/\partial z$ is not used in the nEDM data analysis, it recently improved our understanding of systematic effects that are discussed in Sec. \ref{yk:sec::Systematics}.
	One should also note that it enabled us to measure for the first time the neutron gyromagnetic moment using UCN with a 1 ppm precision \cite{yk:MagneticMomentRatio}.\\
	\\
	The current limitation to lower the $\partial B_0/\partial z$ uncertainty originates from two distinct sources: i) we attribute a 5 pT/cm uncertainty to the CsM data analysis method itself based on simulations and described in Ref \cite{yk:VictorThesis}; ii) an other 5-10 pT/cm coming from CsM offsets, estimated by comparing free spin precession and phase-locked loop measurements during maintenance.\\
		
\subsection{Statistical sensitivity \& systematics budget}
\label{yk:sec::Systematics}
	
	The precise monitoring of $B_0$ down to 0.05 ppm precision ensures our nEDM sensitivity to be only limited by the neutron Larmor frequency estimation, i.e. the neutron statistics (Eq. \ref{yk:eq:ramseysens}).
	The figure of merit of an nEDM experiment follows, in combination of Eq. \ref{yk:eq:principle} and Eq. \ref{yk:eq:ramseysens}:
 	\begin{eqnarray}
	\label{yk:eq:sigmaEDM}
			\sigma_{\dn} = \frac{\hbar}{2 \alpha E T \sqrt{N_{\textrm{cycle}}} \sqrt{N_{\textrm{UCN}}}}
	\end{eqnarray}
	\\
	During the 2013 and 2014 data taking campaigns, on average, we had $\alpha = 0.6$, $E = 10~\textrm{kV/cm}$, $T = 180~\textrm{s}$, $N_{\textrm{UCN}} = 6000 ~ \textrm{/ cycle}$ and 240 cycles per day, leading to a daily sensitivity $\sigma_{\dn} \approx 2.7 \times 10^{-25} \ecm$.
	The 2015 data taking campaign operates with significantly improved performances.
	In particular, we developed a new method to better homogenize the $B_0$ field with the trim coils, and as a result, regularly obtain $\alpha > 0.7$.
	The current UCN source operation provides 2 times more UCN as compared to 2014.
	These improvements enable us to run the experiment with the best ever per day nEDM sensitivity $\sigma_{\dn} \approx 1.3 \times 10^{-25} \ecm$.
	The cumulated statistics already sets our integrated sensitivity to $\sigma_{\dn} \approx 2.5 \times 10^{-26} \ecm$.
	Such a value gives us confidence to be able to improve the current nEDM sensitivity by the end of 2016 in nominal apparatus operation.\linebreak
	\\
	The systematic error budget is given in Table \ref{yk:tab:systematics}.
	In Ref \cite{yk:nEDMlimit}, an exhaustive description of all the effects is given.
	Some of them arise from the $\hg$Hg co-magnetometer.
	One should note the recent direct measurement of the motional false EDM at the origin of the largest systematic effect in our experiment \cite{yk:motionalEDM}.
	This effect, proportional to the vertical magnetic field gradient adds a false nEDM signal of $4.4 \times 10^{-26} \ecm $ in a 10 pT/cm gradient, which is the typical $\partial B_0/\partial z$ uncertainty.
	Fortunately, a dedicated analysis technique described in Ref \cite{yk:nEDMapparatus} enables to precisely determine a working point with zero gradient and therefore circumvent this systematics, at the expense of an other effect: the quadrupole difference.
	As UCN and Hg atoms dynamic are different, both species will not experience the magnetic field inhomogeneities in the same way.
	From this observation, a precise knowledge of the transverse component of $B_0$, called $B_T$, is required, as shown in \cite{yk:MagneticMomentRatio}.
	Extensive magnetic field mappings of the storage volume area have lowered the uncertainty of $B_{\textrm{T}}^2$ to 2 nT$^2$, hence explaining the reduced systematic error budget as compared to the results published in \cite{yk:nEDMlimit}.
	Finally, the precise measurement of $\partial B_0/\partial z$ highlighted a new classe of UCN depolarization, called gravitationally enhanced depolarization \cite{yk:GravDepol} using an innovative UCN Spin Echo technique \cite{yk:SpinEcho}.
	This process influences the motional false EDM correction and must be carefully taken into account in the nEDM analysis.
	In conclusion, the systematic error budget is not restrictive for the current data taking and would already allows us to attain a nEDM sensitivity within the $10^{-27} \ecm$ range.
	
	\begin{table}[t]
	\begin{center}
	\begin{tabular}{c|c}
	\hline  
	Effect &  Status [$\times 10^{-27} \ecm$] \\ 	\hline  
	\hline
	Direct effects & \\
	\hline
	Uncompensated B-Drifts & -0.7 $\pm$ 1.1\\
	Leakage current & 0 $\pm$ 0.05\\
	$v \times E$ UCN & 0 $\pm$ 0.1\\
	Electric forces & 0 $\pm$ 0\\
	Hg EDM & 0.005 $\pm$ 0.015\\
	Hg direct light shift & 0 $\pm$ 0.008\\
	\hline
	Indirect effects & \\
	\hline
	Hg light shift & 0 $\pm$ 0.05\\
	Quadrupole difference & 1.3 $\pm$ 2.4\\
	Dipoles &  0 $\pm$ 3\\
	\hline		
	Total & 0.2 $\pm$ 4.0 \\
	\hline	
	\end{tabular}
	\caption{Systematic errors budget.}
	\label{yk:tab:systematics}
	\end{center}
	\end{table}

\section{The n2EDM phase}
\label{yk:sec:n2EDM}

\begin{figure}
\centering
	\includegraphics[height=3in]{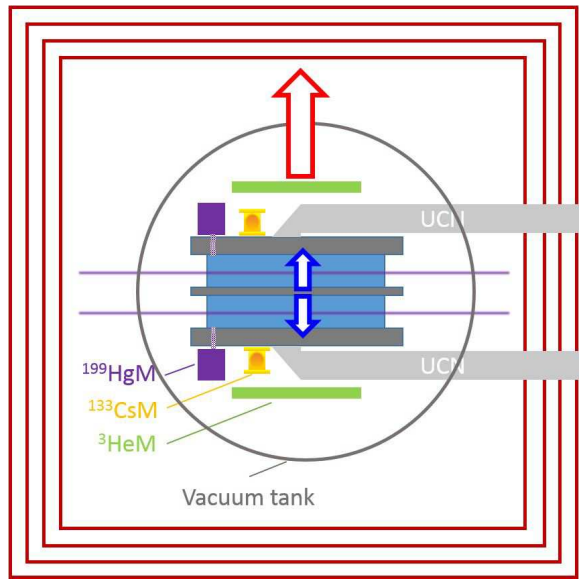}
\caption{Schematic view of a n2EDM spectrometer vertical cut. Red (large) and blue (small) arrows stand for the \textbf{B} and \textbf{E} fields direction respectively.}
\label{yk:fig:n2EDMspectro}
\end{figure}

The next phase at PSI, called n2EDM, is currently in the design and R\&D phase.
With the present UCN source performance, a factor 10 improvement in sensitivity is reachable within a few years of data taking.
This statistical improvement sets our requirements on the n2EDM experiment design (Fig. \ref{yk:fig:n2EDMspectro}) in particular to keep the systematic errors at the same level.
The final goal is to reach the $10^{-28} \ecm$ region.

\subsection{Design}

\paragraph{Magnetic field shield}

	The n2EDM phase will benefit from a cubic multi layers high permeability metal shield having a shielding factor of approximately $10^5$ in the three (x-y-z) directions.
	The homogeneity goal inside the fiducial volume where UCN will be stored is on the 1 pT/cm level, about 20 times better than in the current apparatus.

\paragraph{Precession chamber}

	The main upgrade of this phase lies in a stack of two precession chambers placed on top of each other.
	This new scheme allows for a higher number of neutrons just by volume consideration and simultaneous measurement of both B and E field configurations.
	This configuration will also permit to attain a higher electric field, external electrodes being grounded.
	
\paragraph{Spectrometer optimization}
	
	The current nEDM spectrometer has the disadvantage of being located more than 1 m above the UCN source exit guide and has a much smaller guide diameter leading into the nEDM spectrometer.
	Adapting both features to the UCN source, we expect to increase the number of detected UCN by a factor of up to 5.

\subsection{Research and Development program}
	
	\paragraph{$\hg$Hg co-magnetometry}
	A laser based $\hg$Hg co-magnetometer to correct for B field drift down to 0.01 ppm precision is under study.
	The increased laser light intensity allows a factor 5 improvement in $\hg$Hg frequency precision as compared to the achieved precision with the $^{204}$Hg lamp currently used.
	This improvement is sufficient to handle the neutron sensitivity increase.
	One should also note that a fine tuned laser wavelength will reduce the Hg light shift \cite{yk:MartinThesis}.
	Attention is focused on the laser stability since drifts in air pressure/temperature can cause Hg laser frequency detuning.
	For this purpose, a dedicated laser hut is under construction in the experimental area.

	\paragraph{$\cs$Cs magnetometry}	
	A new type of vector $\cs$Cs magnetometers (VCsM) to better estimate the 3D magnetic field distribution gives already promising results \cite{yk:VCsM}.
	An increased number of such external VCsM as compared to the current array is considered and would further improve the $B_0$ monitoring precision.
	Also, dedicated studies to investigate the CsM offsets issue mentioned in \ref{yk:sec:magnetometry} are under way, one possible solution being the free induction decay (FID) mode \cite{yk:FID}.

	\paragraph{$\he$He magnetometry}
	Continuous effort is pursued to develop $\he$He magnetometers, two designs have been studied: small  $\he$He cells and $\he$He pancakes in gradiometer mode \cite{yk:HeM} \cite{yk:AndreasThesis}.
	$\he$He magnetometers are operated in a free spin precession mode as the HgM and therefore do not suffer from accuracy/offset issue.\\

	\paragraph{$\mathbf{B_0}$ coil}
	A new cubic $B_0$ coil design based on a novel concept is investigated \cite{yk:CrawfordCoil}.
	The idea is to build a $B_0$ coil made of two distinct shells of current loops: an inner part producing the highly homogeneous $B_0$ field; an outer part confining the escaping B field lines to cancel the magnetic field outside the coil, hence preventing for shield magnetization.

\section{Conclusion}
We are currently taking data with an upgraded version of the RAL/Sussex/ILL spectrometer at a high level of performance and will continue for the next years.
With our current sensitivity, we should be in the position to improve the best nEDM limit set in 2006.
Our systematics budget already now offers the possibility to enter in the $10^{-27} \ecm$ sensitivity range, due in particular to extensive magnetic field mapping campaigns and a better understanding of the UCN and $\hg$Hg spin dynamic.
The design of a new apparatus with even higher performance is on the way.


\begin{thebibliography}{99}

\bibitem{yk:Pospelov}
 M. Pospelov and A. Ritz, Ann. Phys. (N. Y). 318, 119 (2005). A. Ritz (this conference)

\bibitem{yk:nEDMlimit}
 J. M. Pendlebury et al arXiv:1509.04411 (2015)

\bibitem{yk:Ramsey}
 N. F. Ramsey, Phys. Rev. 78, 695 (1950).
 
\bibitem{yk:GreenHgM}
 K. Green et al Nucl. Instruments Methods Phys. Res. Sect. A Accel. Spectrometers, Detect. Assoc. Equip. (1998).

\bibitem{yk:nEDMapparatus}
 C. A. Baker et al Nucl. Instruments Methods Phys. Res. Sect. A Accel. Spectrometers, Detect. Assoc. Equip. 736, 184 (2014).

\bibitem{yk:UCNsourceNIMA}
H. Becker et al. Nucl. Instruments Methods Phys. Res. Sect. A Accel. Spectrometers, Detect. Assoc. Equip. 777, 20 (2015).

\bibitem{yk:UCNsource}
B. Lauss Phys. Procedia 51, 98 (2014).

\bibitem{yk:USSA}
S. Afach et al arXiv:1502.06876 (2015).

\bibitem{yk:NANOSC}
 G. Ban et al Nucl. Instruments Methods Phys. Res. Sect. A Accel. Spectrometers, Detect. Assoc. Equip. 611, 280 (2009).

\bibitem{yk:CsMarray}
 P. Knowles et Nucl. Instruments Methods Phys. Res. Sect. A Accel. Spectrometers, Detect. Assoc. Equip. 611, 306 (2009).

\bibitem{yk:MagneticMomentRatio}
 S. Afach et al Phys. Lett. B 739, 128 (2014).
 
 \bibitem{yk:VictorThesis}
V. Helaine PhD thesis, Universite de Caen/Basse-Normandie (2014).

\bibitem{yk:motionalEDM}
S. Afach et al Eur. Phys. J. D 69, 225 (2015).

\bibitem{yk:GravDepol}
S. Afach et al Phys. Rev. D 92, 052008 (2015).

\bibitem{yk:SpinEcho}
S. Afach et al arXiv:1506.00446 (2015).

\bibitem{yk:MartinThesis}
M. C. Fertl PhD thesis, ETH Zurich (2013).

\bibitem{yk:VCsM}
 S. Afach et al Opt. Express 23, 22108 (2015).

\bibitem{yk:FID}
Z. D. Grujic‡, P. A. Koss, G. Bison, and A. Weis, Eur. Phys. J. D 69, 135 (2015).

\bibitem{yk:HeM}
H.-C. Koch et al Eur. Phys. J. D 69, 202 (2015).

\bibitem{yk:AndreasThesis}
A. Kraft PhD thesis, Johannes Gutenberg-Universitat (2012).

\bibitem{yk:CrawfordCoil}
C. Crawford Workshop on "Challenges of the world-wide experimental search for the electric dipole moment of the neutron", Ascona, Switzerland (2014)


\end{thebibliography}
\end{document}